\documentclass[
reprint, 
superscriptaddress,
 amsmath,amssymb,
 aps, physrev,
]{revtex4-2}

\usepackage{graphicx}
\usepackage{dcolumn}
\usepackage{bm}
\usepackage{color}
\usepackage{makecell}
\usepackage{multirow}
\usepackage{comment}
\usepackage{booktabs}
\usepackage{tabularx} 

\begin{document}

\preprint{APS/123-QED}

\title{CelloAI: Leveraging Large Language Models for HPC Software Development in High Energy Physics
}%

\author{Mohammad Atif}
\email{fmohammad@bnl.gov}
\affiliation{
Brookhaven National Laboratory, NY (USA)
}
\author{Kriti Chopra}
\affiliation{
Brookhaven National Laboratory, NY (USA)
}
\author{Ozgur Kilic}
\affiliation{
Brookhaven National Laboratory, NY (USA)
}
\author{Tianle Wang}
\affiliation{
Brookhaven National Laboratory, NY (USA)
}
\author{Zhihua Dong}
\affiliation{
Brookhaven National Laboratory, NY (USA)
}
\author{Charles Leggett}
\affiliation{
Lawrence Berkeley National Laboratory, CA, (USA)
}
\author{Meifeng Lin}
\affiliation{
Brookhaven National Laboratory, NY (USA)
}
\author{Paolo Calafiura}
\affiliation{
Lawrence Berkeley National Laboratory, CA, (USA)
}
\author{Salman Habib}
\affiliation{
Argonne National Laboratory, IL (USA)
}

\collaboration{HEP-CCE}

\date{\today}

\begin{abstract}

Next-generation High Energy Physics (HEP) experiments will generate unprecedented data volumes, necessitating High Performance Computing (HPC) integration alongside traditional high-throughput computing. However, HPC adoption in HEP is hindered by the challenge of porting legacy software to heterogeneous architectures and the sparse documentation of these complex scientific codebases.
We present CelloAI, a locally hosted  coding assistant that leverages Large Language Models (LLMs) with retrieval-augmented generation (RAG) to support HEP code documentation and generation.  This local deployment ensures data privacy, eliminates recurring costs and provides access to  large context windows without external dependencies. CelloAI addresses two primary use cases, code documentation and code generation, through specialized components. For code documentation, the assistant provides:  (a) Doxygen style comment generation for all functions and classes by retrieving relevant information from RAG sources (papers, posters, presentations), (b) file-level summary generation, and (c) an interactive chatbot for code comprehension queries. For code generation, CelloAI employs syntax-aware chunking strategies that preserve syntactic boundaries during embedding, improving retrieval accuracy in large codebases.  The system integrates callgraph knowledge to maintain dependency awareness during code modifications and provides AI-generated suggestions for performance optimization and accurate refactoring. We evaluate CelloAI using real-world HEP applications from ATLAS, CMS, and DUNE experiments, comparing different embedding models for code retrieval effectiveness.  Our results demonstrate the AI assistant's capability to enhance code understanding and support reliable code generation while maintaining the transparency and safety requirements essential for scientific computing environments. 

\end{abstract}

\maketitle

\section{Introduction}

Recent advancements in large language models (LLMs) have paved the way for tools that can enhance the software development process for scientists. LLMs excel at two key tasks: code documentation in natural language and code generation in a given programming language~\cite{minaee2024large,gu2025effectiveness}. However, commercially available tools are often restricted by limited context window sizes, encounter usage limits, or incur substantial costs for large-scale development and modernization of extensive scientific codebases. Moreover, data privacy and security concerns~\cite{Yao2024} present additional barriers to adoption. Thus, a programmatic framework that can address some of these deficiencies and can be executed on local infrastructure is needed. This framework should guarantee long context windows for large codebases and must be capable of utilizing the latest open models offline for code generation. 
They must also be adept at handling edge cases, synchronization issues, performance, and have sufficient understanding of architectural differences.
The key idea is to develop a privacy-preserving, locally-executed framework that leverages LLMs with advanced prompt engineering to automatically generate contextually-aware documentation and code for scientific software. 
This approach addresses the unique challenges of complex research codebases while eliminating external dependencies and enhancing code understanding across scientific domains.

In this context, we have developed \textbf{CelloAI}, an LLM-based code assistant specifically designed for software documentation, understanding and code generation in high performance computing (HPC) environments with a focus on  high energy physics (HEP) applications. This specialized focus addresses the growing  need within the  HEP community to leverage HPC resources for processing the dramatically increased data volumes  anticipated from  next-generation experiments, such as the High-Luminosity Large Hadron Collider (HL-LHC) experiments~\cite{ciangottini2025analysis} and the future Deep Underground Neutrino Experiments (DUNE)~\cite{lopez2025dune}. Scientific HPC software development presents unique challenges that distinguish it from general-purpose programming and render standard LLM-based coding assistants inadequate.  The highly specialized nature of HPC software results in limited training data availability, creating knowledge gaps in existing models.  Furthermore, the stringent correctness requirements of scientific research conflict  with the inherently probabilistic nature of  LLM  outputs, where even minor errors can invalidate computational results.  Additionally,  HEP software ecosystems feature domain-specific frameworks and complex inter-dependencies that demand sophisticated  understanding of the code logic and architectural relationships. To address these challenges, CelloAI employs three complementary mechanisms: 1) Retrieval-Augmented Generation (RAG) enhanced with HEP-specific data sources (source code, documentation, tutorials, and publications, etc.); 2) intelligent code chunking strategies that preserve the semantic meaning during embedding; and 3) the integration of callgraph knowledge to capture comprehensive codebase relationships and dependencies.

\section{Background and Related Work}

Large language models (LLMs) have rapidly transformed code generation and analysis. 
This transformation began with the pioneering work by \citet{chen2021evaluating} which introduced Codex and released the evaluation set HumanEval as a benchmark for code generation.
Codex was fine-tuned on public GitHub repositories, and solved 28.8\% of functional programming tasks on HumanEval with only 12 billion parameters. This breakthrough spurred the development of open foundation models specializing in code and established the foundation for today's diverse ecosystem of AI-powered coding tools.

The success of Codex inspired major contributors across the field to develop their own approaches. Google released AlphaCode 2 for novel programming solutions using specialized sampling and filtering techniques \cite{li2022competition}. Meta's Llama-based code models \cite{roziere2023code}, ranging from 7 billion to 70 billion parameters, achieved up to 65\% scores on HumanEval benchmarks through instruction-tuning. IBM's Granite Code Models  \cite{granite2024granite} span 3 billion to 34 billion parameters across 116 programming languages, while the BigCode community released StarCoderBase and StarCoder—15.5 billion parameter models trained on permissively licensed code  \cite{li2023starcoder}. 

Building on these foundational developments, the landscape of AI for code generation has evolved into four distinct categories:

\begin{enumerate}
    \item \textbf{Proprietary Models}: These foundation LLMs are the most popular closed-weight models for remote inference. They do not offer options for on-premise deployment and their usage is metered by tokens and policy limits. They typically get upgraded quickly as the vendors constantly upgrade the back-ends. Examples include OpenAI GPT-o3/5, Claude-Opus, Gemini 2.5 Pro.
    
    \item \textbf{API-based Services}: These are command-line or SaaS layers that sit on top of a proprietary model and orchestrate multi-step ``agent” workflows for code generation. They offer access to agentic loops to fix bugs, add tests, and run shell commands. Their usage is also charged per token. Examples include APIs from OpenAI, Gemini, Anthropic, Perplexity AI, and Together AI. 
        
    \item \textbf{AI Assistants plugged into IDE or Shell}: These are lightweight clients that live inside an IDE or Linux shell. They feed local code contexts to one or more remote or local models. The choice of model is often configurable by the user. They are tightly integrated with IDE/shell (inline completions, refactor, run tests) but offer the possibility of offline work when paired with local LLMs. Examples include Gemini CLI, aider, Windsurf, Claude Code, Github Copilot, and Cursor.
    
    \item \textbf{Open-weight LLMs}: Several organizations release model checkpoints under permissive licenses, but they typically require custom workflows and computational infrastructure. Open-weight models democratize research and on-premise deployment. Their unique advantage is full data-privacy, offline generation, and a transparent access to large context windows. Examples include Meta's Llama-3.3, IBM's Granite-20b-code, OpenAI gpt-oss-120b, typically downloaded from the organization's Hugging Face \cite{huggingface} repository.

\end{enumerate}

While the aforementioned general-purpose coding assistants excel at many tasks such as creating code from scratch based on instructions, helping with debugging codes, writing unit tests etc., they face significant limitations when applied to  high-performance computing (HPC) and scientific applications. These domains present unique challenges that expose fundamental gaps in current LLM capabilities.  Scientific software often involves unpublished research or export-controlled knowledge making on-premise deployment not just preferable but necessary for compliance with  institutional policy, national security rules, or grant agreements.
Beyond privacy concerns, scientific workflows demand bit-wise reproducibility which conflicts with the stochastic nature of the LLMs. Since these state-of-the-art models are updated regularly, the reproducibility that research requires is compromised.  This problem is compounded by the high-stake nature of scientific computing, where codes control multi-million dollar experimental facilities such as colliders/accelerators or beam-line magnets, where LLM hallucinations can have catastrophic consequences.

Beyond these operational concerns, LLMs face fundamental technical limitations in scientific computing.  The massive size of scientific codes necessitate larger context windows and specialized fine-tuning. More critically, modifying or adapting large code bases requires understanding the inter- and intra-functional dependencies which these LLMs might miss, potentially leading to broken functionality or subtle bugs that propagate through complex systems. This dependency blindness is exacerbated by the documentation problem in scientific computing, where novel research is regularly published but corresponding codebases often lack sufficient details, creating fragmented knowledge that LLMs (and humans) struggle to piece together coherently. 
Recently studies have also demonstrated that all the top open- and closed-weight LLMs exhibit significantly lower performance in multi-turn conversations than single-turn, with an average drop of 39\% across six generation tasks \cite{laban2025llms}. These demand safety guardrails in terms of stringent checks, static analysis and rigorous testing of generated code. 


Recognizing these limitations, the research community has developed specialized frameworks that directly address scientific computing needs.  For example, PaperCoder transforms machine-learning papers into executable repository-level code via a multi-stage pipeline (planning, analysis, generation) and evaluates outputs using both reference-based and reference-free protocols, achieving strong replication scores on newly introduced PaperBench benchmarks \cite{seo2025paper2code}. HiPerRAG \cite{gokdemir2025hiperrag} is a scalable Retrieval-Augmented Generation framework that combines a high-throughput multimodal PDF parser with a query-aware encoder to index and retrieve context from over 3.6 million scientific articles, achieving high accuracy on benchmarks. 
Complementing these approaches, other researchers have created domain-specific chatbots by pairing large language models with embedding based retrieval from scientific PDFs, delivering valid answers with reduced hallucinations \cite{yager2023domain}. 

\noindent \textbf{Performance portability and optimization}

\noindent Despite remarkable progress, tailoring LLM-based development to HPC code generation remains challenging~\cite{Godoy2024, Chen2024}. 
HPC-Coder, fine-tuned on a novel dataset of parallel programs, can auto-complete HPC kernels, insert OpenMP pragmas, and predict performance trends—capabilities beyond what generic LLMs achieve \cite{nichols2024hpc}. In the domain of performance-portable libraries, ChatBLAS uses LLMs to generate a portable, AI-produced BLAS implementation for CPU and GPU targets, matching or surpassing vendor benchmarks for level-1 routines \cite{valero2024chatblas}. Most recently, LASSI (“LLM-based Automated Self-correcting pipeline for Scientific codes”) automates bi-directional translation between OpenMP and CUDA scientific codes \cite{dearing2024lassi}. By incorporating self-correcting loops—feeding compilation and runtime errors back into the model — LASSI achieves up to 85 \% successful translations, with 78 \% of OpenMP to CUDA and 62 \% of CUDA to OpenMP runs executing within 10 \% of original performance.
Yet these specialized systems reveal limitations -- while excelling at CUDA and OpenMP, they struggle with newer portable abstractions such as HIP or Kokkos \cite{godoy2023evaluation}, highlighting the ongoing challenge of keeping pace with the evolving HPC technologies. 

To summarize, while general-purpose coding LLMs enable rapid prototyping even for exascale software, real-world HPC adoption requires addressing fundamental issues of  privacy,  predictable cost envelopes, reproducibility, workflow-specific UX, and model transparency.
The most effective solutions combine open-weight and fine-tuning for sensitive kernels with stringent guardrails and version control ensuring that scientific results remain trustworthy while researchers can still benefit from the productivity gains that have driven the current LLM renaissance in software development.

\begin{figure}
    \centering
    \includegraphics[width=0.75\linewidth]{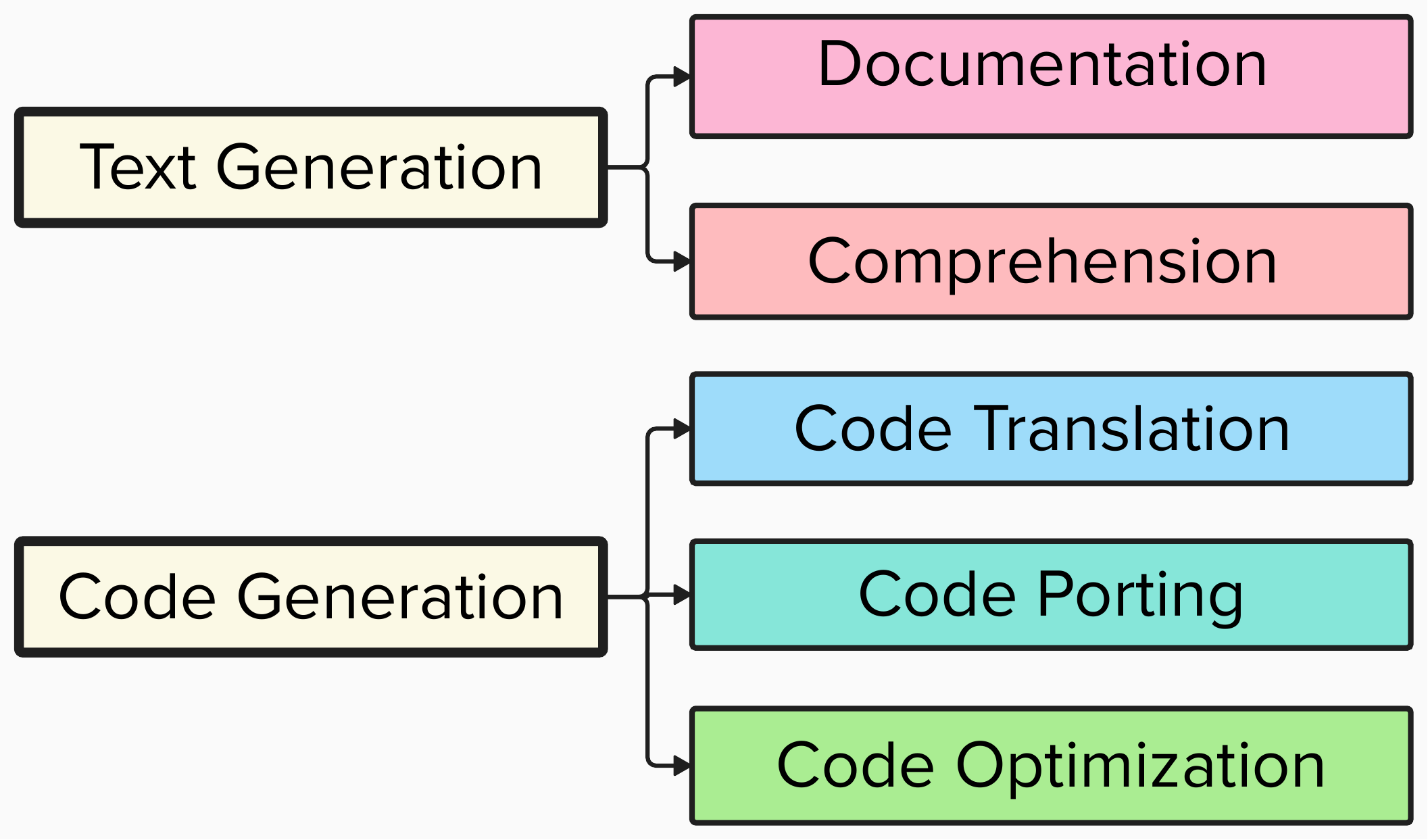}
    \caption{Categorization of an coding assistant's tasks into AI generation. }
    \label{fig:ai_tasks}
\end{figure}

\section{Motivation and Baseline Evaluation}

A useful AI assistant for scientific computing must address several distinct but interconnected challenges spanning both text and code generation. 
Text generation capabilities are fundamentally for two key scenarios: (a) Creating documentation from diverse text sources such as manuscripts, slide decks etc. and (b) Providing an interactive code comprehension support through a chatbot like interface to on-board new researchers or developers. 
However, code generation presents significantly greater complexity and risk as any hallucination or unchecked AI generated code can have disastrous consequences in scientific computing environments. 
The challenge intensifies when considering  the dynamic nature of scientific software development, where performance requirements drive continuous evolution across hardware architectures.  
Scientific applications increasingly require code translating  between programming models to leverage novel hardware capabilities \cite{atif2024porting,lin2023portable}. 
For example, most real world applications now include GPU code originally written for CPUs but subsequently translated or adapted to CUDA. 
The emergence of AMD and Intel GPUs alongside NVIDIA has made performance portability a critical consideration in scientific computing.  

AI assistants must therefore excel at cross-platform code translation while preserving correctness and performance characteristics. 
Beyond translation, these systems must tackle open ended optimization tasks that require identifying performance bottlenecks and proposing algorithmic improvements. 
These challenges demand deep understanding of both code structure and computational principles.
Figure \ref{fig:ai_tasks} shows the above discussed tasks for an AI coding assistant.
In this study, we consider applications identified by High Energy Physics Center for Computational Excellence (HEP-CCE) as testbeds for performance portability \cite{atif2023evaluating}. 



\subsection{Testbeds}

\begin{table*}[ht]
\centering
\caption{Summary of the four scientific computing testbeds discussed in this study. }
\label{tab:testbed_summary}
\begin{tabularx}{\textwidth}{@{}ll>{\raggedright\arraybackslash}X>{\raggedright\arraybackslash}X>{\raggedright\arraybackslash}X@{}}
\toprule
\textbf{Testbed} & \textbf{Experiment} & \textbf{Primary Purpose} & \textbf{Key Computational Task(s)} & \textbf{Implementation/Portability Notes} \\
\midrule
\textbf{FastCaloSim} & ATLAS & Fast calorimeter simulation & Parametrized simulation of electromagnetic/hadronic showers & Reduces Geant4 simulation time. Ports and containers exist for various compilers and hardware \cite{atif2023evaluating,atif2025packaging}. \\
\addlinespace
\textbf{P2R} & CMS & Track reconstruction mini-app & Track state propagation \& Kalman filter updates & Lightweight app implemented with various portability technologies for CPUs and GPUs \cite{ather2024exploring}. \\
\addlinespace
\textbf{Patatrack} & CMS & Pixel track reconstruction & Standalone Patatrack pixel tracking; track reconstruction & Extracted from the official CMS software (CMSSW). Several ports for various backends exist \cite{bocci2020heterogeneous}. \\
\addlinespace
\textbf{WireCell} & DUNE & LArTPC simulation \& reconstruction & Signal simulation (rasterization, scatter-add, FFT) & Modern C++ with data flow paradigm. Supports multi-threaded CPU; GPU portability evaluated \cite{lin2023portable,dong2023evaluation}. \\
\bottomrule
\end{tabularx}
\end{table*}

\begin{table*}[]
    \centering
    \begin{tabular}{c|c|c|c|c}
     & \multicolumn{2}{c}{\textbf{Code}} &  \multicolumn{2}{|c}{\textbf{Text}} \\
       \hline
     & No. of Files  & Lines of Code &  No. of Files & No. of Words \\
     \hline
     FastCaloSim           & 247  & 26,100 & 15 & 341,587 \\
     \hline
     P2R                   & 29   & 23,957 & 14 & 302,192 \\
     \hline
     Patatrack            & 1269 & 124,948 & 14 & 302,192 \\
     \hline
     WireCell Kokkos       & 22   & 3,083 & 22 & 1,525,551 \\
     \hline
     WireCell OpenMP       & 16   & 2,162 & 22 & 1,525,551 \\
     \hline
    \end{tabular}
    \caption{Characteristics of selected codebase and their associated text corpora.}
    \label{tab:codebase_char}
\end{table*}


We select testbeds from HEP experiments (ATLAS, CMS, DUNE) typically used for simulations and track reconstruction.
These applications are FastCaloSim, Patatrack, P2R, and WireCell and are listed in Table \ref{tab:testbed_summary} along with their characteristics.
They encompass a diverse range of programming models including CUDA, HIP, OpenMP, Kokkos, SYCL, Alpaka, and std::par.
We aim to automate code translation from one programming model to another to either (a) enhance portability (e.g., CUDA to OpenMP/Kokkos), or, (b) to capitalize on vendor libraries for performance (e.g., OpenMP to CUDA).
Table \ref{tab:codebase_char}  lists quantitative    characteristics such as the number of files, lines of code and size of their associated text corpora.
The code metrics were calculated using the Count Lines of Code library \cite{adanial_cloc}.  All the PDF files were  converted to markdown format using {marker-pdf} \cite{marker_pdf} before ingestion into database. 

\subsection{Prompt}

\begin{table}[]
    \centering
    \begin{tabular}{c|c|c}
    \multicolumn{3}{l}{
    \makecell{
\fbox{%
\parbox{0.45\textwidth}{%
You are an expert high-performance computing software developer. I have a BASE$\_$IMPL codebase called **NAME** for High-Energy Physics simulations. I need you to: 
1. **Scan the entire codebase** and identify every function implemented for GPU execution, marked with IDENTIFIER. 
\\
2. **Output a flat list** of those function names under the heading Table 1, one per line, along with the source file path where each appears. 
\\
First, **only produce the list of BASE$\_$IMPL functions**. After you’ve confirmed the list, devise a strategy to replace each function with an equivalent PORT$\_$IMPL version considering important operations like data transfer, memory initialization, floating point operations, and reduction. Do not write any code.
}%
}
    }
    }  \\
     \textbf{BASE$\_$IMPL} & \textbf{IDENTIFIER} & \textbf{PORT$\_$IMPL} \\
     \midrule
     CUDA   & $\_\_$global$\_\_$, $\_\_$device$\_\_$ & OpenMP\\
     \midrule
     Kokkos & Kokkos::parallel$\_$for                & CUDA\\
     \midrule
     OpenMP & pragma omp target                      & CUDA\\
     \midrule
    \end{tabular}
    \caption{Prompt for collecting GPU kernels. The keywords BASE$\_$IMPL, IDENTIFIER, PORT$\_$IMPL are replaced in for different programming models. }
    \label{tab:code_keywords}
\end{table}

While analyzing and upgrading large scientific code bases with LLMs, it is important to have transparency of the context and its contents.
The first step in porting an application from one programming model to another involves identifying and assembling the relevant kernels.
Although this appears straightforward, different AI solutions exhibit vastly different performance characteristics.
To illustrate this variability,  we  compared the performance of industrial solutions like Google Gemini, OpenAI ChatGPT, and Github Copilot.
It should be noted that the above solutions are not RAG-based approaches in the strict sense, but employ different mechanisms of scanning source code.
Rather than engineering optimal prompts for each solution we adopt a common prompt (see Table \ref{tab:code_keywords}) to enable a fair comparison.
This prompt requests the AI solution to collect all relevant GPU kernels identified by specific key words and present them  in a table format.

\begin{table}[]
    \centering
    \begin{tabular}{c|c|c|c|c}
     & \textbf{Total} & \makecell{\textbf{gemini-cli}\\gemini-2.5-pro} & \makecell{\textbf{ChatGPT}\\o3} & \makecell{\textbf{Copilot}\\Sonnet4.0}  \\
     \hline
     \textbf{CUDA}        & & & \\
     FastCaloSim          & 8    & 8       & 8    & 8 \\
     Patatrack           & 1490 & -*      & 1267 & 121 \\
     P2R                  & 119  &  82     & 108  & 84 \\
     \hline
     \textbf{Kokkos}      & & & \\
     FastCaloSim          & 2    & -*   & 2  & 2 \\
     Patatrack           & 169  & -*   & 77 & 51 \\
     P2R                  & 14   & 7    & 6  & 6  \\
     WireCell             & 76   & 39   & 16 & 21\\
     \hline
     \textbf{OpenMP}      & & & \\
     FastCaloSim          & 5    & 4**     & 4 & 4 \\
     Patatrack           & 25   & -*      & 2 & 7 \\
     WireCell             & 68   & 11      & 16 & 6 \\ 
     \hline
    \end{tabular}
    \caption{List of GPU kernels collected by various AI assistants: terminal-based gemini-cli, web browser-based ChatGPT o3, and IDE-based Github Copilot. Gemini-cli either loops through files or concatenates the entire codebase before collecting patterns. * Error: the input token count exceeds the maximum number of tokens allowed (1048576). ** Error: hallucinates function names. 
    }
    \label{tab:aisol_recall}
\end{table}

Table \ref{tab:aisol_recall} lists the number of kernels recalled from different HEP applications. We observe that the coverage is highly uneven across programming models and that recall performance degrades  as the kernel count increases.
All three assistants perform well on small CUDA and OpenMP kernel sets in FastCaloSim, yet their performance deteriorates significantly for the same directives in Patatrack and WireCell,  sometimes inventing non-existing kernel names. 
Gemini-cli -- a command line interface to Gemini -- attempts to read and concatenate all  files in the code repository, thus exceeding the one million tokens context window limit  for Patatrack. 
This inconsistency forces developers to  manually cross‑check results, negating any productivity gains. 
For the Patatrack CUDA workload (1490 kernels) ChatGPT o3 retrieves 1267 kernels  whereas Copilot surfaces only 121. 
A similar trend appears for Kokkos kernels: ChatGPT o3 finds 77/169 (46\%) in Patatrack and only 16/76 (21\%) in WireCell; Copilot hovers around 30\%. 
Thus, we find that none of the commercial tools can sustain high recall once the search space exceeds a few hundred symbols.
These findings reinforce the need for a transparent, retrieval‑centered workflow for code generation and  rationalize the development of a tailored AI solution based on a deterministic  enumeration of every relevant kernel.  Without such capabilities, migration of a large HEP application from CUDA to HIP or from OpenMP to Kokkos becomes unreliable.


\section{Design and Implementation}


CelloAI employs a specialized retrieval mechanism designed to address the unique challenges of scientific code comprehension and generation. The system's effectiveness depends critically on assembling contextually relevant code fragments and explanatory text before the language model inference begins. This section describes the architectural components that ensure the context provided to the LLM is both comprehensive and pertinent to the user query. 

\subsection{Separate Collections and Configurable Embedding Models for Code and Text}

Collections are the fundamental unit of storage and query offered by ChromaDB for compartmentalizing vector embeddings \cite{chromadb}.
CelloAI optimizes retrieval performance by embedding code and text through separate collections. 
This separation is essential for queries that require both algorithmic understanding and contextual explanations, such as, 
 ``Explain the underlying algorithm for the function X as it relates to Y".  
Source code and scientific documentation exhibit fundamentally different characteristics within an information‑retrieval pipeline. Code is structured, syntactic and follows programming language conventions, whereas scientific text is semantic, descriptive and domain-specific. Recognizing these differences,  CelloAI treats them as independent data streams from the initial harvesting stage through final retrieval stage. The system processes each corpus separately:  
code and text are chunked using appropriate strategies and embedded with distinct, user‑selectable models into a ChromaDB. 
Each  chunk, whether code or prose, is also augmented with its file path metadata to preserve provenance and enable precise source attribution. 
This decoupled design enables rapid experimentation with new open‑weight embedding models, and ensures independent tuning of the retrieval quality.

\subsection{Syntax-aware Code Chunking}

Common RAG pipelines rely on text-centric chunking schemes like the RecursiveCharacterTextSplitter,  which partition documents into fixed-length windows or heuristically sized passages based on whitespace and punctuation. 
While adequate for prose, these approaches are not optimal for specialized applications or source code \cite{yepes2024financial}, where there can be critical syntactic and semantic boundaries. 
In C++ codebases, naive chunking strategies create several problems. 
For example, a sliding window may bisect a template parameter list, or orphan a `catch' block from its corresponding `try' statement, or strip away enclosing `namespace' and `using' directives. 
Consequently, retrieved snippets frequently lack complete function signature or necessary context, leading to incorrect code generation when fragments from unrelated functions are inappropriately combined. 
To address this limitation, CelloAI implements a Tree-sitter driven chunking strategy.
Tree-sitter is a parser generator tool and a parsing library used to build a concrete syntax tree for a given source file \cite{treesitter} across multiple programming languages.
A Treesitter-driven chunker 
emits self-contained units like complete function definitions, method implementations, or class definitions. 
This strategy removes code fragmentation and thus yields higher accuracy in code retrieval for downstream generation tasks.

\subsection{CelloRetriever: Unified Code and Text Retrieval with Pattern Matching}

\begin{figure}
    \centering
    \includegraphics[width=0.95\linewidth]{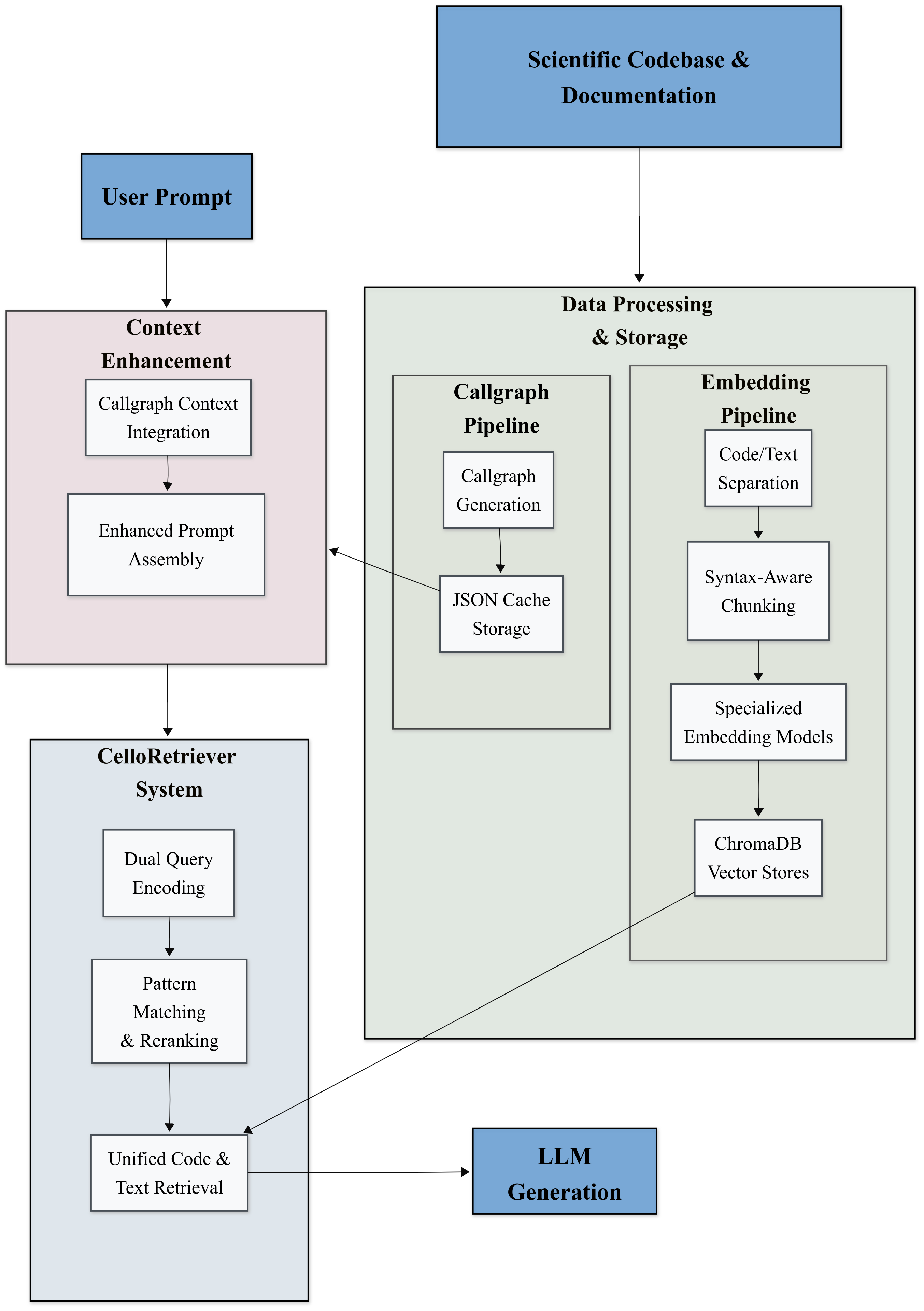}
    \caption{CelloRetriever: Combined retrieval pipeline for code and text coupled with pattern matching and callgraph prompt enhancement.}
    \label{fig:celloretriever}
\end{figure}

CelloRetriever integrates the retrieved source code chunks and  scientific text into a unified context  for the language model. 
Every user query is encoded twice -- once with the code embedding model and once with the text embedding model. 
A configurable number of top-ranked documents are retrieved independently from each vector database.
To optimize the final context composition, CelloRetriever applies a lightweight pattern‑matching pass that identifies strings within \`{}\`{}\`{}triple backquotes\`{}\`{}\`{} and then re‑orders these candidates. 
This hybrid ranking has three practical advantages: (a) Balanced context: the code snippets provide concrete algorithmic implementation, while text snippets carry the scientific rationale, giving the LLM both “how” and “why” within a single prompt, (b) Higher precision: the pattern matching layer reranks the documents with exact symbol matches at the top, thus rewarding semantically dissimilar but algorithmically relevant chunks, and (c)
Low computational overhead: the additional scoring introduces minimal overhead, as retrieval operation constitute a small  fraction of the total computational cost compared to language model generation.
Figure \ref{fig:celloretriever} depicts the workflow of CelloRetriever.

\subsection{Automatic Prompt Enhancement with Callgraphs}

Large scientific codebases often contain critical dependencies spanning  multiple layers of function calls, making it difficult for developers to fully understand the impact of potential modifications. 
To prevent an LLM from suggesting changes that could silently break downstream logic, CelloAI automatically enhances every prompt with contextual dependency information derived from  callgraphs.
This feature can be controlled with a configuration switch ENHANCE$\_$PROMPT$\_$WITH$\_$LINEAGE.
We construct these graphs once during ingestion with Doxygen (C/C++/CUDA/HIP) and cache them as JSON.
When CelloRetriever identifies a function pertinent to the user's query, we append a two‑hop lineage, i.e., its immediate callers and callees summarized in natural language into the prompt context.
This simple addition delivers the following benefits: (a) Dependency awareness: the LLM gains visibility into how a target function integrates within the broader execution flow, thus reducing likelihood of suggestions that would compromise downstream code, and (b) Enhanced Explainability: the lineage information provides users with a quick recap of the affected call chain without manually traversing multiple files, improving development efficiency and comprehension. 
Thus, by explicitly incorporating  structural dependencies into the generation context, callgraph‑aware prompting transforms the LLM from an isolated code generator into a dependency aware collaborator. This architectural decision substantially improves  both the safety and practical utility of the generated code suggestions in complex scientific computing environments.

\section{Experiments and Results}

We evaluate CelloAI's effectiveness across its core functionalities using real-world HEP applications from ATLAS, CMS,  and DUNE experiments.  Our evaluations focus on three key areas: CelloRetriever's performance and context assembly (including separate collections,  syntax-aware chunking and embedding models comparisons), automated document generation capabilities, and code translation performance using various open-weight language models. Each experiment demonstrates how CelloAI's specialized components address the unique challenges of scientific code understanding and generation. 

\subsection{Retrieval of Code/Text for Context}
We evaluate CelloRetriever's effectiveness by examining each of its key design components. 

\subsubsection{Separate Collections for Code and Text}
\begin{table}[h]
    \centering
    \begin{tabular}{c|c|c}
     & \multicolumn{2}{c}{Retrieved [top 50]}\\
     \hline
     & Code & Text \\
     \hline
     FastCaloSim (CUDA)  & 25 & 25 \\
     \hline
     P2R (CUDA)          & 42 & 8 \\
     \hline
     Patatrack (CUDA)   & 50 & 0 \\
     \hline
     WireCell (Kokkos)   & 22 & 28 \\
     \hline
     WireCell (OpenMP)   & 32 & 18 \\
     \hline
    \end{tabular}
    \caption{Retrieved code and text from the same collection.}
    \label{tab:same_collection}
\end{table}

To demonstrate the necessity of separate collections, we first examine retrieval behavior when the code and text share a single vector store.
Using the prompt in Table \ref{tab:code_keywords} with Lajavaness/bilingual-embedding-large embedding model, we analyze the distribution of source types in retrieved contexts. 
Table \ref{tab:same_collection} shows the code-to-text split among the top fifty retrieved documents from  unified collection. The results reveal significant imbalance that varies with the repository characteristics.  For instance, in Patatrack, all fifty retrieved documents consist exclusively of the source code, likely due to the repository's substantial codebase size  (see Table \ref{tab:codebase_char}).
CelloAI addresses this limitation through independent collections and allowing the user to configure the retrieval quotas from each source type.  This design ensures transparent and predictable context assembly, providing users an explicit control over the material fed to the language model rather than solely relying on embedding scores. The configurable approach enables balanced contexts that combine algorithmic details from code with conceptual explanations from documentation, essential for effective scientific code assistance.
\subsubsection{Syntax-aware Code Chunking}
\begin{table}
    \centering
    \begin{tabular}{c|c|c}
     & \multicolumn{2}{c}{\makecell{Routines in Retrieved \\Code [top 40]}}\\
     \hline
     & Partial & Complete \\
     \hline
     FastCaloSim (CUDA)      & 21 & 12 \\
     \hline
     P2R    (CUDA)           & 20 & 19 \\
     \hline
     Patatrack  (CUDA)      & 12 & 6 \\
     \hline
     WireCell (Kokkos)   & 33 & 4 \\
     \hline
     WireCell (OpenMP)   & 31 & 3 \\
     \hline
    \end{tabular}
    \caption{Retrieved codes from separate collections: number of partial and complete routines in 40 top ranked chunks. Notice that the sum of routines is less than 40 as (i) there are chunks containing only header includes, and (ii) lengthy routines cover multiple chunks.}
    \label{tab:natural_chunk}
\end{table}
To demonstrate the importance of boundary-aware chunking for scientific code retrieval, we compare CelloAI's syntax-aware approach against conventional fixed window chunking. We analyze code fragmentation by counting incomplete function retrievals in the assembled context. 
Table  \ref{tab:natural_chunk} quantifies  how many of the first 40 code chunks returned by the chunking strategy are ``partial routines,” i.e., fragments that begin or end in the middle of a function body.  The baseline fixed-window approach produces substantial fragmentation across all tested applications, though the extent varies by codebase characteristics. 
This fragmentation significantly degrades context quality by introducing incomplete function signatures, unmatched scope delimiters, and missing variable declarations which in turn leads to poor LLM output quality.  CelloAI's syntax-aware chunking approach circumvents this fragmentation through the Tree-sitter-based method, hence providing complete routines to the language models for accurate code understanding and generation.

\subsubsection{Benchmarking Embedding Models for Code and Text}


\begin{table*}
    \centering
    \begin{tabular}{c|c||c|c||c|c||c|c||c|c}
     & & \multicolumn{2}{c||}{\makecell{ sentence-\\transformers/\\all-MiniLM-L6-v2} } & \multicolumn{2}{c||}{\makecell{microsoft/\\codebert-base} } &\multicolumn{2}{c||}{ \makecell{Lajavaness/\\bilingual-\\embedding-large} } & \multicolumn{2}{c}{\makecell{ Salesforce/\\SFR-Embedding\\-Code-2B$\_$R} }    \\
     \hline
     & Params & \multicolumn{2}{c||}{22.7M} & \multicolumn{2}{c||}{125M} & \multicolumn{2}{c||}{560M} & \multicolumn{2}{c}{2.61B}    \\
     \hline
     & \makecell{Embedding\\Dimensions} & \multicolumn{2}{c||}{384} & \multicolumn{2}{c||}{768} & \multicolumn{2}{c||}{1024} & \multicolumn{2}{c}{2304}   \\
     \hline
     & \makecell{Initial\\Release} & \multicolumn{2}{c||}{August 2021} & \multicolumn{2}{c||}{July 2020} & \multicolumn{2}{c||}{June 2024} & \multicolumn{2}{c}{January 2025}   \\
     \hline
     \hline
     & Total & Base &  \makecell{Cello\\Retriever} & Base &  \makecell{Cello\\Retriever} & Base &  \makecell{Cello\\Retriever} & Base &  \makecell{Cello\\Retriever}   \\
     \hline
     \textbf{CUDA}       & & & & & & & & &  \\
     FastCaloSim         & 8    & 4  & 8   & 0 & 8  & 4   & 8   & 6  & 8    \\
     Patatrack          & 1490 & 50 & 171 & 1 & 127 & 134 & 191 & 40 & 179  \\
     P2R                 & 119  & 4  & 53  & 6 & 53 & 11  & 44  & 9  & 53   \\
     \hline
     Score               & & 0.189	& 0.520 & 0.017 & 0.510 & 0.227	& \textbf{0.790 }& 0.284 & 0.359   \\
     \hline
     \hline
     \textbf{Kokkos}     &      & & & & & & & &   \\
     FastCaloSim         & 2    & 2 & 2 & 0 & 2 & 1 & 2 & 0 & 2 \\
     Patatrack          & 169  & 43 & 44 & 0 & 36 & 43 & 44 & 1 & 44  \\
     P2R                 & 15   & 9 & 15 & 1 & 15 & 11 & 15 & 11 & 15    \\
     WireCell            & 76   & 24 & 32 & 8 & 32 & 36  & 40  & 28 & 40   \\
     \hline
     Score               & & 0.543 &	0.670 &	0.043 &	0.658 &	0.490 &	\textbf{0.697} &	0.277 &	\textbf{0.697}  \\
     \hline
     \hline
     \textbf{OpenMP}     & & & & & & & & &  \\
     FastCaloSim         & 5    & 0 & 5 & 0 & 5 & 3 & 5 & 2 & 5 \\
     Patatrack          & 25   & 0  & 25 & 0 & \textbf{1} & 1 & 25 & 0 & 25 \\
     WireCell            & 68   & 52 & 55 & 54 & 55  & 54  & 55  & 54  & 55  \\ 
     \hline
     Score               & & 0.255 &	\textbf{0.936} &	0.265 &	0.616 &	0.478 &	\textbf{0.936} &	0.398 &	\textbf{0.936}  \\
    \end{tabular}
    \caption{Quantifying the completeness of retrieved code: comparison of various embedding models across applications and parallel programming paradigms.}
    \label{tab:text_vs_code_embed}
\end{table*}

We evaluate CelloRetriever's performance across four distinct embedding models that vary in specialization,  release date, embedding dimensions, and total number of parameters.
The selected models include two text based models (sentence-transformers/all-MiniLM-L6-v2 and Lajavaness/bilingual-embedding-large) and two code-specialized embedding models (microsoft/codebert-base and Salesforce/SFR-Embedding-Code-2B$\_$R). To quantify retrieval effectiveness, we develop a completeness metric based on kernel identification tasks. Using the prompt in Table \ref{tab:code_keywords}, which requests constructing a table of GPU kernels, we count the number of retrieved kernels for each embedding model. For each programming paradigm, we measure the fraction of correctly retrieved code blocks against a curated ground-truth list from various applications. 
The final score is calculated as the mean of completeness ratio across various applications, where a score of 1.0  indicates  perfect retrieval of all relevant code blocks. 

The scores presented in Table \ref{tab:text_vs_code_embed} demonstrate the following: (a) text-based embedding models perform as well as or better than code specialized embedding models for text-heavy prompt, (b) the newer and larger embedding models typically achieve higher scores, and (c) CelloRetriever's pattern matching consistently  enhances retrieval across all embedding models and programming paradigms (CUDA, Kokkos,  OpenMP ).  
For instance, with the CUDA dataset, the Lajavaness/bilingual-embedding-large model's score increased from a baseline of 0.227 to 0.790 with  CelloRetriever. Similarly, for OpenMP tasks, the custom retriever enables the newer models to achieve a near-perfect score of 0.936, representing significant improvement over baseline performances. 
These results demonstrate CelloRetriever's robustness in  overcoming standard retrieval limitations regardless of the underlying embedding model characteristics.

\begin{table*}[]
    \centering
    \begin{tabular}{c|c|c|c|c|c|c|c|c}
    \textbf{LLM} & & \textbf{Size} & & \textbf{Coverage} & \multicolumn{3}{c|}{\textbf{Compilable/Correct}} & \makecell{\textbf{Instruction}\\\textbf{Adherence}} \\
     & & Billion (B) & & & Easy & Moderate & Hard & \\
    \hline
    \multirow{2}{*}{\makecell{agentica-org/\\DeepCoder-14B-Preview}} 
    & \multirow{2}{*}{\makecell{Code Gen}} & \multirow{2}{*}{\makecell{14B}}
    & Base      & 0/8 & N/- & N/- & N/- & \multirow{2}{*}{5.5/10}  \\
    & & & CelloAI & 8/8 & Y/Y & N/- & N/- &  \\
    \hline
    \multirow{2}{*}{\makecell{microsoft/\\Phi-4-reasoning-plus}} 
    & \multirow{2}{*}{\makecell{Reasoning}} & \multirow{2}{*}{\makecell{14B}} 
    & Base      & 4/8 & N/- & N/- & N/- & \multirow{2}{*}{6/10} \\
    & & & CelloAI & 8/8 & N/- & Y/Y & N/- &  \\
    \hline
    \multirow{2}{*}{\makecell{openai/gpt-oss-20b}} 
    & \multirow{2}{*}{\makecell{Text Gen}} & \multirow{2}{*}{\makecell{20B}}
    & Base      & 4/8 & Y/Y  & N/-  & N/- & \multirow{2}{*}{7/10} \\
    & & & CelloAI & 8/8 & Y*/Y & Y*/N & N/- & \\  
    \hline
    \multirow{2}{*}{\makecell{Qwen/Qwen2.5-\\Coder-32B-Instruct}} 
    & \multirow{2}{*}{\makecell{Code Gen}} & \multirow{2}{*}{\makecell{32B}} 
    & Base      & 4/8 & Y*/Y & N/- & N/- & \multirow{2}{*}{8/10} \\
    & & & CelloAI & 8/8 & Y*/Y & Y/N & N/-  & \\    
    \hline
    \multirow{2}{*}{\makecell{meta-llama/\\Llama-3.3-70B-Instruct}} 
    & \multirow{2}{*}{\makecell{Text Gen}} & \multirow{2}{*}{\makecell{70B}}
    &  Base     & 4/8 & Y/Y  & N/- & N/- & \multirow{2}{*}{7/10} \\
    & & & CelloAI & 6/8 & Y*/Y & N/- & N/- & \\
    \hline
    \multirow{2}{*}{\makecell{openai/gpt-oss-120b}} 
    & \multirow{2}{*}{\makecell{Text Gen}} & \multirow{2}{*}{\makecell{120B}}
    & Base      & 4/8 &  Y/Y  & N/-  & N/- & \multirow{2}{*}{8/10} \\
    & & & CelloAI & 8/8 & Y/Y & Y*/N & Y/N & \\  
    \hline
    \multirow{2}{*}{\makecell{mistralai/Mistral-\\Large-Instruct-2411}} 
    & \multirow{2}{*}{\makecell{Text Gen}} & \multirow{2}{*}{\makecell{123B}}
    & Base      & 2/8 & Y/Y & N/- & N/- & \multirow{2}{*}{6/10} \\
    & & & CelloAI & 8/8 & Y/Y & Y/Y & N/- & \\
    \hline    
    ChatGPT-o3 & Browser &  & & 8/8 & Y/Y & N/- & N/- & \multirow{2}{*}{3/10} \\  
    ChatGPT-5  & &  & & 8/8 & N/- & N/- & N/- &     
    \end{tabular}
    \caption{Comparative evaluation of various LLMs for porting FastCaloSim kernels from CUDA to OpenMP. Coverage measures how many kernels were ported, * indicates minor issues that were easily fixable.}
    \label{tab:ai_bench}
\end{table*}

\begin{table}[]
    \centering
    \begin{tabular}{c|c|c}
    \multicolumn{3}{l}{
    \makecell{
\fbox{%
\parbox{0.45\textwidth}{%
You are an expert GPU and high-performance computing software engineer. You will convert a set of BASE$\_$IMPL IDENTIFIER kernels from the FastCaloSim project into equivalent, performant PORT$\_$IMPL kernels using the above strategy. The resulting code must compile, target new GPU architectures, and integrate cleanly into the existing CMake build. 
\\
1. For every BASE$\_$IMPL function in Table 1 rewrite the implementation in PORT$\_$IMPL. 
\\
2. Preserve numerical results bit-wise where feasible (double precision). 
\\
3. Insert explicit device memory management where the BASE$\_$IMPL runtime previously handled mapping: allocate once per event batch and reuse. 
\\
4. Use features from PORT$\_$IMPL that are known to be more efficient. 
\\
5. Supply a two-sentence performance note per kernel (expected speed-up, occupancy bottlenecks).
}%
}
    }
    }  \\
    \end{tabular}
    \caption{Prompt for porting GPU kernels. The keywords BASE$\_$IMPL, IDENTIFIER, PORT$\_$IMPL are same as Table \ref{tab:code_keywords}.}
    \label{tab:port_prompt}
\end{table}

\subsection{Code Documentation and Summarization}

CelloAI provides an out-of-the-box utility for code documentation and summarization that generates Doxygen style comments and file level summaries. This functionality addresses the common problem of sparse documentation in research software by leveraging both code structure analysis and contextual information from scientific literature. The documentation process operates through systematic repository traversal using Tree-sitter to identify spans of each top-level function, method, or class.  
For each identified code element, the system combines the function signature with relevant contextual information retrieved from text databases and streams this enriched context to the local LLM. The language model is then prompted to generate a Doxygen-style comment block containing a brief summaries, parameter descriptions and return value documentation.  
To ensure safety and maintain code integrity, CelloAI implements the following guardrail.  All generated documentation is wrapped in a comment block (/** … */)  before reinsertion into the source code, rendering any hallucinated code snippets inert and preventing execution.   Each generated comment includes a watermark statement  identifying it as LLM-generated and prompting reviewers to validate the information.  
The system preserves original code formatting during documentation insertion process ensuring that downstream compilers or static analyzers see only syntactically correct, annotated code. This automated approach significantly reduces the manual effort required for documentation while maintaining the scientific rigor necessary for research software development.

\subsection{Interactive Coding Assistant}

Finally, we provide an interactive coding assistant that runs locally from the command line.
This can be used to prompt LLMs to perform concrete tasks such as enumerating kernels, explaining call chains, proposing refactors, or translating code. 
The chatbot retains conversation history in memory so the follow-up prompts refine the same working context rather than starting from scratch.
We evaluate several open weight LLMs encompassing parameter size (14 billion to 123 billion parameters) and specializations (code generation, text generation, reasoning)  with a two-turn test on FastCaloSim.
Here, we follow-up the previous prompt for collecting kernels (Table \ref{tab:code_keywords}) with a second prompt (see Table \ref{tab:port_prompt}) to actually port the collected kernels from CUDA to OpenMP.
The first prompt primes the context by collecting relevant code chunks and their callgraph neighborhoods.
The second prompt instructs the LLM to write OpenMP code for GPU target offloading.
The ability of an LLM is measured by its ability to write `\#pragma omp target' regions and insert explicit `target enter/exit data' mappings or `map(tofrom:...)' for the device buffers.
The prompt also requests a short explanation or rationale for the generated code.

It is evident from Table \ref{tab:ai_bench} that adding CelloAI’s retrieval consistently increases coverage across all on-premise models.
Note that most baseline models are able to identify and generate code for 4/8 kernels, with the exception of Mistral-Large-Instruct-2411 (2/8).
However, with CelloAI's pipeline each model ports all kernels, except Llama-3.3-70B-Instruct which ports 7/8 kernels despite all of them being available in the context.
Additionally, due to the augmented caller information from callgraphs, Qwen2.5-Coder-32B, Phi-4-reasoning-plus, openai/gpt-oss-20b, openai/gpt-oss-120b, and Mistral-Large-Instruct-2411 are able to generate the CPU function that calls the GPU kernels.
Qwen2.5-Coder-32B is noted for its unique ability to write multiple `\#pragma omp target data map' clauses in this CPU function -- a very important requirement for correctly porting codes to OpenMP.
{OpenAI's o3, gpt-oss-20B, and gpt-oss-120B write relevant code for device memory allocation/deletion and movement using `omp$\_$target$\_$alloc/free/memcpy' with the 120B parameter model being the only one that writes all the three.}

Next, we evaluate the generated OpenMP target kernels for their compilability and correctness. 
The eight kernels are divided into three levels of difficulty: easy, moderate, and hard.
The easy kernels are straightforward tests that print a message from several GPU threads or initialize a GPU array.
The moderate kernels involve recognizing patterns for atomic operations or have more than one simple operation.
The hard kernels are computationally intensive that simulate events in FastCaloSim.
We notice that most local LLMs are able to write easy kernels correctly, however, only a few are able to correctly port moderate kernels.
None of the models we tested have yet been able to port the hard kernels owing to the incorrect or missing host to device memory mappings.
Phi-4-reasoning-plus demonstrates mixed performance. While it successfully handles the moderate kernel (omp target atomic) and provides verbose explanations, it  fails on simpler kernels by inventing nonexistent functions and produces poorly formatted, unreadable code. 
OpenAI gpt-oss-20B consistently produced well-structured responses, often organizing information in tables and demonstrating correct multi-clause mapping directives.
However, we observe a drop in performance with ChatGPT-5 (as compared with ChatGPT-o3) as it fails to correctly implement even the straightforward kernels.  
A common error across several models involves inappropriate use of  `\#pragma omp target teams distribute parallel for' directive for operations that require no loops, such as updating a single value on GPU memory.  In these cases, a simple `\#pragma omp target' directive suffices, and the unnecessary loop construct causes compilation errors.
Qwen2.5-Coder-32B-Instruct, Mistral-Large-Instruct-2411, gpt-oss-20B, and gpt-os-120B circumvent this issue by implementing the for loop with conditional $i < 1 $, using a single team and single thread to achieve functionally correct code.
{OpenAI's gpt-oss-120B is the only model among the ones we test which produces hard kernels that compile. 
However they throw runtime errors as it is unable to connect `is$\_$device$\_$ptr' with `omp$\_$target$\_$alloc', instead incorrectly using `map' clauses.} 

We also evaluate instruction adherence of the models via a blinded human study.
The model outputs were anonymized and shuffled, then scored independently by the authors on a 0–10 scale based on the adherence to the prompt’s required elements (formatting, inclusion of requested code/artifacts, explanatory rationale, and the specific ``performance notes” field). Scores were averaged per sample and then per model to produce a single instruction-adherence rating. As CelloAI’s retrieval system changes the input context but does not influence instruction adherence, we report a common average per model in Table \ref{tab:ai_bench}. 
We observe that ChatGPT scored lower primarily for omitting the requested performance notes alongside the kernels, however, other models tend to meet the formatting and checklist requirements more consistently.
It should also be noted that these models differ vastly in size and active parameters, thus fair comparison requires more investigation. 

{Despite promising results, the framework comes with clear limitations and failure modes. 
While retrieval is reproducible, LLM generation has high stochasticity that needs to be tamed.
As the testing and development progresses, the decoding hyperparameters such as temperature, top-p, top-k, and tail-free sampling will need to be tuned.
On the one hand, lowering the temperature can enhance reproducibility yet it risks locking in incorrect or sub-optimal kernels. 
On the other hand, higher temperature explores alternatives but can yield entirely different code when the initial code contains bugs. 
Additionally, benchmarking of LLM-generated content is an open problem.
For example selecting the best model among code generation/text generation/reasoning models demands a standardized, end-to-end framework that compiles, executes, and compares numeric correctness, memory behavior, and performance across target architectures. 
Instruction adherence is also a failure mode as some models forget or ignore constraints in long contexts and requires novel mitigation strategies.
}

\section{Outlook}

To summarize, CelloAI raised coverage for most open-weight models.
The presence of callgraph in the context helped several models to emit the CPU wrappers that invoke the GPU kernels. 
However, there are limitations -- while ``easy” kernels compile reliably, ``moderate” ones succeed only for a subset of models, and “hard” kernels fail primarily due to incorrect or missing memory map clauses. We also observed workarounds that avoid compile errors but obscure data-motion requirements. 
These findings indicate that while retrieval cures coverage issues, the data-movement reasoning and directive selection remain the bottlenecks for correctness at scale.
These problems might require fine-tuning the models leading to specialized porting models.

In the future, we will add inline comment generation at edit points so the assistant inserts concise, and granular comments and explanations.
CelloAI might also serve as a community training playground with permissive HEP kernel pairs to train the HEP community about inner working of LLMs.
Technically, our next steps focus on stronger guardrails (static analyzers and unit-test stubs co-generated with code) to ensure correctness. 
Together, these enhancements aim to close the gap on ``hard” kernels while keeping CelloAI usable, auditable, and fully on-premise.
{Hence, CelloAI is a first step toward an agentic framework that (i) ports GPU kernels across backends, (ii) writes unit tests from physics specifications and reference datasets, and (iii) executes and benchmarks while feeding compiler diagnostics, runtime errors, and performance metrics so that developing optimal kernels is possible while the stochasticity of generation is constrained.}

\begin{acknowledgements}
This work is supported by US Department of Energy, Office of Science, Office of High Energy Physics under the High Energy Physics Center for Computational Excellence (HEP-CCE), a collaboration between Argonne National Laboratory, Brookhaven National Laboratory, Fermilab, Oak Ridge National Laboratory, and Lawrence Berkeley National Laboratory. 
\end{acknowledgements}


\bibliography{apssamp, LLM4HPC}






























\end{document}